\documentclass[11pt,a4paper]{article}
\usepackage{amsmath}
\usepackage{enumerate}
\usepackage{amssymb}
\usepackage{graphicx}
\usepackage{multirow}
\usepackage{xcolor}
\usepackage{bm}
\usepackage[normalem]{ulem}

\newcommand{\paczynski}{Paczy\'nski}
\usepackage{jcappub}

\begin{document}

\title{The local and distant Universe:  stellar ages  and $H_0$}

\author[1,2]{Raul Jimenez,}
\author[3,4]{Andrea Cimatti,}
\author[1,2]{Licia Verde,}
\author[3,5]{Michele Moresco,}
\author[6,7,8]{Benjamin Wandelt}

\affiliation[1]{ICCUB, University of Barcelona, Marti i Franques 1, Barcelona, 08028, Spain.}
\affiliation[2]{ICREA, Pg. Lluis Companys 23, Barcelona, 08010, Spain.} 
\affiliation[3]{Dipartimento di Fisica e Astronomia, Universita di Bologna, Via Gobetti 93/2, I-40129, Bologna, Italy.}
\affiliation[4]{INAF - Osservatorio Astrofisico di Arcetri, Largo E. Fermi 5, I-50125 Firenze, Italy.}
\affiliation[5]{INAF - Osservatorio di Astrofisica e Scienza dello Spazio - Istituto Nazionale di Astrofisica, via Gobetti 93/3, I-40129, Bologna, Italy.}
\affiliation[6]{Sorbonne Universit\'e, CNRS, UMR 7095, Institut d'Astrophysique de Paris, 98 bis bd Arago, 75014 Paris, France.}
\affiliation[7]{Sorbonne Universit\'e, Institut  Lagrange  de  Paris  (ILP),  98  bis bd Arago, 75014 Paris, France.}
\affiliation[8]{Center for Computational Astrophysics, Flatiron Institute, 162 5th Avenue, 10010, New York, NY, USA.}

\emailAdd{raul.jimenez@icc.ub.edu; a.cimatti@unibo.it; liciaverde@icc.ub.edu; michele.moresco@unibo.it; bwandelt@iap.fr}

\abstract{The ages of the oldest stellar objects in our galaxy provide an independent test of the current cosmological model as they give a lower limit to the age of the Universe. Recent accurate parallaxes by the Gaia space mission, accurate measurements of the metallicity of stars, via individual elemental abundances, and advances in the modelling of stellar evolution, provide new, higher-precision age estimates of the oldest stellar populations in the galaxy: globular clusters and  very-low-metallicity stars. The constraints on the age of the Universe, $t_U$, so obtained are determined from the local Universe and at late time. It is well known that local and early-Universe determinations of another cosmological parameter closely related to the age of the Universe, the Hubble constant $H_0$, show a $\gtrsim 3 \sigma$ tension. In the standard cosmological model, $\Lambda$CDM,  $t_U$ and $H_0$ are related by the matter density parameter $\Omega_{m,0}$. 
We propose to combine local $t_U$ constraints with late-time $\Omega_{m,0}$  estimates in a $\Lambda$CDM framework, to obtain a low-redshift $H_0$ determination that does not rely on early Universe physics.
A proof-of-principle of this approach with current data gives $H_0=71\pm2.8$  ($H_0= 69.3 \pm 2.7$) km s$^{-1}$ Mpc$^{-1}$ from globular clusters (very-low-metallicity stars) with excellent prospects for improved constraints in the near future.}

\maketitle

\section{Introduction}

The standard model of cosmology ($\Lambda$CDM) has been overwhelmingly successful. Yet, our poor understanding of dark energy and lack of direct evidence for particle dark matter,  constitute a motivation for exploring options beyond the standard model.
One possible hint that there might be physics beyond the standard $\Lambda$CDM is the disagreement at the $3-4 \sigma$ level between the Universe expansion rate at $z=0$, $H_0$, as directly measured by the local cosmic  distance ladder (the latest determination is presented in Ref.~\cite{Riess+18}) and the one inferred by Cosmic Microwave Background (CMB) observations provided by the Planck space mission (the latest results presented in Ref.~\cite{Planck18} --Planck18--). This tension was first quantified in Ref.~\cite{Tension} and has stimulated extensive discussions (e.g., \cite{Efstathiou,Bernal,2017NatAs,Feeney} and references therein).  It is important to note that CMB data provide an inferred value for $H_0$ which is derived within a  well-specified cosmological model, typically a flat $\Lambda$CDM model and relies on standard assumptions about early-Universe physics.  While the local measurement is ultimately based on astrophysics of stars, the CMB-based determination  depends on the cosmological model. The two results are fully  independent from each other and highly complementary.  In fact, thanks to observations such as Baryon Acoustic Oscillations (BAO) and cosmological Type 1A Supernovae (SNe), the cosmic distance ladder  (an thus the expansion history of the Universe) has been mapped and linked from $z=0$ (local $H_0$ determination) to $z=1100$ (CMB) via the combination of BAO and SNe (e.g., Ref.~\cite{Cuesta,Aubourg}).
 
Both teams (Planck18 and Riess et al.) and the wider scientific community, have done a series of very thorough analyses to uncover and eliminate any source of systematic uncertainties and found none. Before invoking new physics, it is important to explore other probes that can provide a  measurement of the current expansion rate $H_0$, and in a broader context, set the early Universe against the late Universe. 

There is evidence that the $H_0$ tension cannot easily be solved by changing the expansion history of the Universe between $z=0$ and $z=1100$, as observations constrain $H(z)/H_0$ to have the  redshift dependence,  the ``shape", predicted by a standard  $\Lambda$CDM model e.g., Ref.~\cite{Bernal}. It is the normalisation of this quantity that does not match. The $H_0$ tension is therefore an ``anchor" problem. The type of unknown or unaccounted for physics to be invoked  depends on the anchor to be adjusted.  Should the late-time anchor be endorsed (i.e., confirmed through  different  lines of evidence), then the physics in the two decades of expansion before  recombination would need to be modified e.g., Ref.~\cite{sounds_discordant}.  Should the high-redshift (early time) anchor be endorsed, then very local and low redshift physics  would need to be altered.

The CMB-based determination has been re-derived independently of the Planck data \cite{Addison} and broadly confirmed. 
The Hubble constant can be determined from the late-time Universe, with no (or minimal) cosmological assumptions in several different ways that we will now discuss.

The standard cosmic distance ladder determination has provided so far the most stringent constraints, and has been scrutinized, cross-checked and tested extensively. While it relies on stellar and astrophysical modelling (e.g., Cepheids, Supernovae), and on accurate measurements of parallaxes,  it is cosmology-independent (besides the assumption of the local space-time metric to be Euclidian).

Another approach is provided by the so-called cosmic chronometers~\cite{GP} which use the ages of stellar populations to obtain $H_0$ by fitting the cosmic chronometer \cite{CC1,CC2,CC3,CC4,CC5,CC6} data at $z>0$. 
By using  Gaussian processes  to extrapolate the cosmic chronometer data to $z=0$, Ref.~\cite{GP} obtained $H_0 = 68.5 \pm 3.5$ km s$^{-1}$ Mpc$^{-1}$ and $\Omega_{m,0} = 0.32 \pm 0.05$\footnote{Hereafter $\Omega_{m,0}$ denotes the matter density at $z=0$.} without relying on CMB data.
 Recently, Ref.~\cite{morescoGP} exploited the possibility of extending the standard Gaussian process approach (Multi-Task Gaussian Process, MTGP) to be able to combine coherently different probes, taking advantage of the complementarity of different probes in redshift and in parameter space. To this purpose, several late-Universe probes were considered, namely cosmic chronometers \cite{CC1,CC2,CC3,CC4,CC5,CC6}, Supernovae Type-Ia (SNe) \cite{SN}, and  BAO \cite{BAO1, BAO2, BAO3, BAO4, BAO5, BAO6}. This method reconstructs the Hubble parameter $H(z)$ without any cosmological assumption, hence providing a model-independent estimate of $H_0$. They found $H_0 = 68.5 \pm 2.9$ km s$^{-1}$ Mpc$^{-1}$. The  still somewhat large error bar of these determinations, and the fact that  value of $H_0$ found is in between that of Planck18 and the local distance ladder,  implies that these measurements are not yet able to shed light on the current discrepancy for $H_0$.
 
Other approaches also do not clearly resolve the $H_0$ controversy. Time-delay lensing distance measurements  provide a low-redshift $H_0$ determination, $H_0=72.5^{+2.1}_{-2.3}$ km s${-1}$ Mpc$^{-1}$, \cite{Birrer}. 
 The recent discovery of electromagnetic counterparts to gravitational waves has facilitated the measurement of the Hubble constant in a purely geometric way. While the current number of objects is very low, and thus the uncertainty on $H_0$ somewhat large, they are nevertheless very interesting already. Ref.~\cite{GW} reports $H_0 = 70^{+12}_{-8}$ km s$^{-1}$ Mpc$^{-1}$.

 In analogy to $H_0$, the current age of the Universe, $t_U$,is a parameter of the $\Lambda$CDM model \cite{Knox} that can also be measured both through early-Universe and late Universe observations. It is predicted by Planck18 within the $\Lambda$CDM model to be $t_U=13.8 \pm 0.02$ Gyr because of the very accurate measurement of the position of the first acoustic peak in the CMB temperature power spectrum. The very high accuracy (0.1\%) of this prediction (see Ref.~\cite{Knox} for its physical origin) makes $t_U$ an interesting probe to compare the early and late Universe. This comparison can be achieved by estimating the cosmological-model independent value derived from the oldest stellar objects  in the Galaxy at $z=0$.

A different approach is to use the absolute ages of stellar objects at $z=0$ \cite{JimenezGC,JimenezTreu} in combination with CMB-independent estimates of $\Omega_{m,0}$  e.g., from galaxy clustering,  to constrain $H_0$. This approach must assume a cosmological model (e.g., a flat $\Lambda$CDM), but it  is a late-time measurement and  does not rely on  assumptions about the early-Universe physics. In the past, given the large uncertainties in the ages of the oldest stellar objects, and lack of sufficiently precise low-redshift determinations of $\Omega_{m,0}$, this approach was not very useful.

Recently,  more accurate determinations of $\Omega_{m,0}$ have become available via measurements of higher-order clustering of galaxies in the BOSS sample \cite{Bis1,Bis2,Bis3} and by the application of the Alcock-\paczynski\ test to voids e.g.,\cite{voidAP,voids}.  These methods do not rely on  the CMB as they use only clustering measurements at the redshift of observation. 
Determinations of the age of the oldest stars have also recently undergone a transformational evolution.
 With the recent data release  by the Gaia satellite of accurate parallaxes, it is now  possible to compute more accurate distances to the oldest galactic globular clusters, thus removing the main error budget in determining globular cluster ages when using the main-sequence fitting method (e.g., Ref.~\cite{omalley}).

The purpose of this paper is to explore what current new data on stellar ages and on low redshift determinations of $\Omega_{m,0}$,  can tell us about  the early-vs-late Universe tension both in terms of estimate of the age of the Universe and  the inferred value of $H_0$. We present a proof-of-principle approach  using state-of-the art  stellar ages determinations, of comparing low-redshift measurements of $H_0$ and $t_U$ with inferred values from CMB observations. While current  determinations of $t_U$ from the local Universe, combined  with late-time  matter density estimates,  might  yield an $H_0$ constraint that does not yet have the robustness and accuracy to match  local $H_0$ determinations, forthcoming data will greatly improve the situation.

\section{Methods}

Determination of stellar ages in the nineties provided one of the first hints that the dominant cosmological model at the time (an Einstein-de-Sitter Universe) needed  revision (see e.g., Ref.~\cite{ostriker,JimenezGC,Spinrad}). Old stellar populations were determined to be older than $1/H_0$, the age of the Universe in that model (see e.g. Ref.~\cite{JimenezGC}). Of course, the age of stellar objects at $z=0$ is just a lower limit to the age of the Universe and, by itself, does not constrain the cosmological model, as changes in $H_0$ and $\Omega_{m,0}$ can accommodate an Einstein-de-Sitter Universe. 

In the past, in order to break this degeneracy,  a determination the stellar ages of the oldest galaxies at $z \gg 0$ proved crucial. This was first achieved by Ref.~\cite{Dunlop}. It is revealing to see Fig.~18 in Ref.~\cite{Spinrad} which shows the exclusion of the  Einstein-de-Sitter Universe once the ages of GCs are taken into account.  
This philosophy has been further developed in the cosmic chronometer method, with the first truly cosmological-model-independent determination of the redshift evolution of the Hubble parameter, $H(z)$ \cite{CC2,CC3,CC5,CC6}.

\subsection{Age of the Universe estimate from stellar ages}
\label{sec:methods:ages}
\begin{figure}
 \begin{centering}
\includegraphics[width=.48\columnwidth]{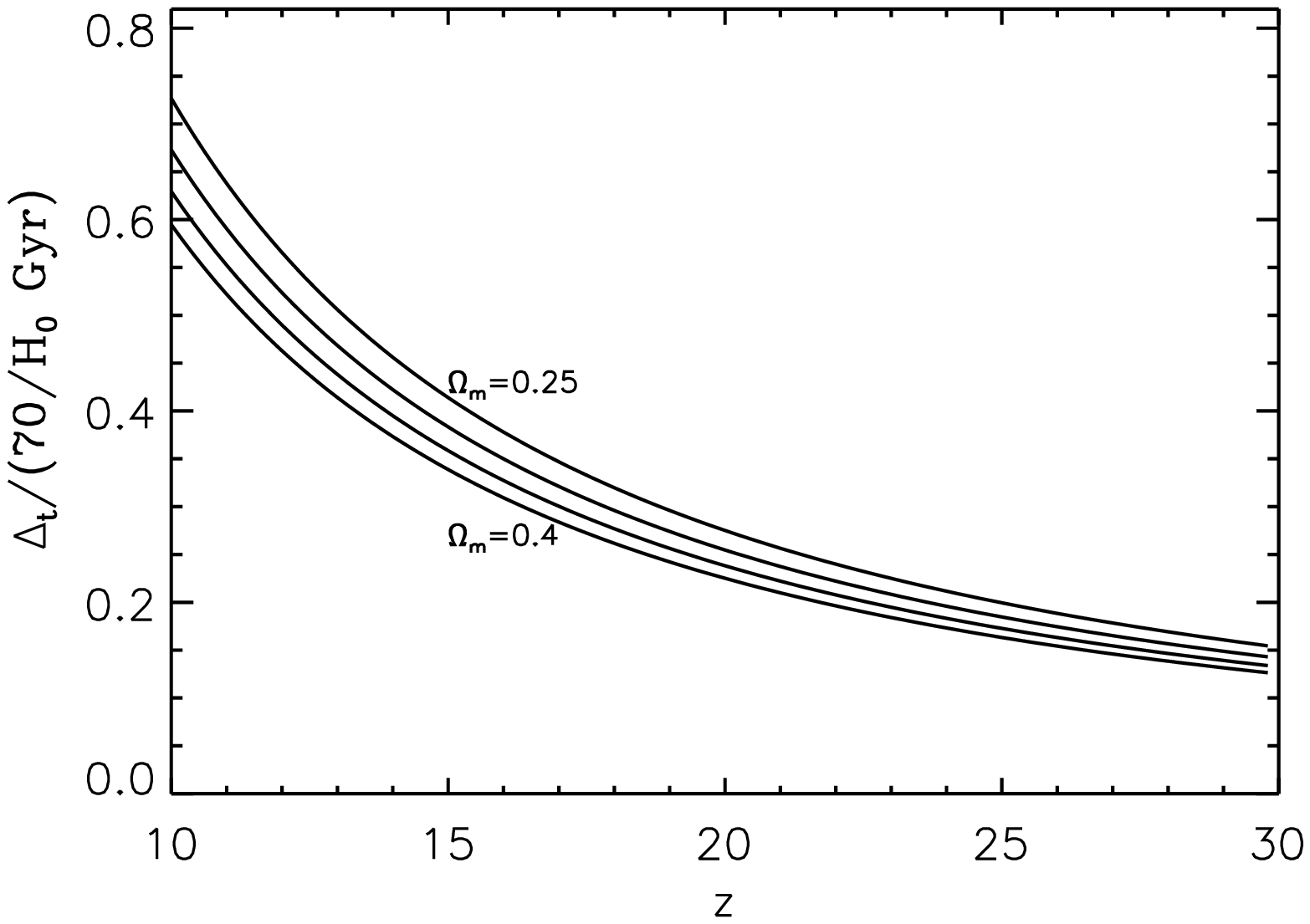}
\includegraphics[width=.48\columnwidth]{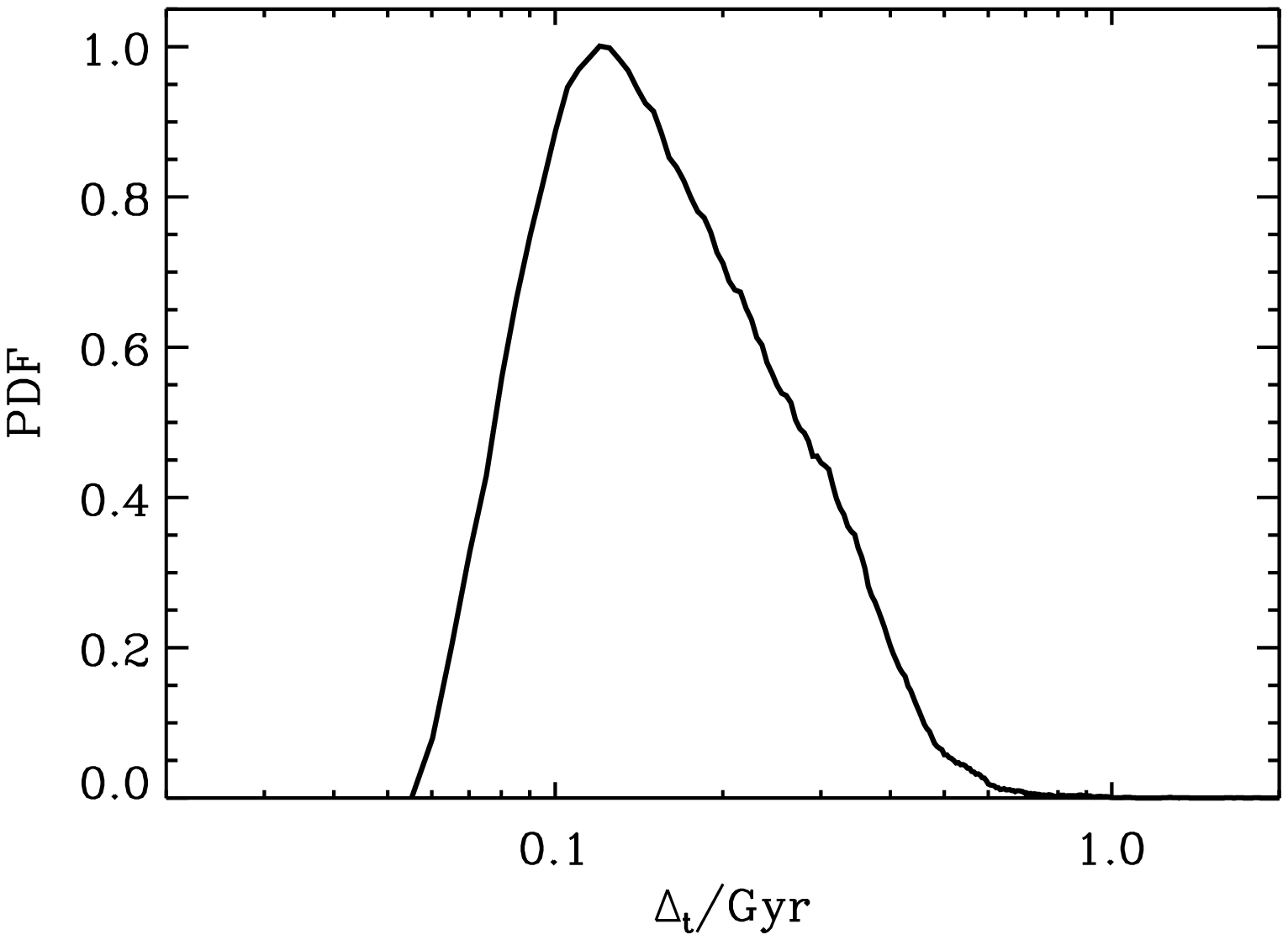}
\caption{Left panel: $\Delta_t$ as a function of redshift of formation of the oldest stellar objects. The current farthest known confirmed galaxy is at $z=11.1$  \cite{Oesch} , therefore we argue that  the oldest stars must have formed at  least at this redshift, but also demonstrate in the text that assuming it to be $z \sim 8$ does not change our conclusions. Note that for $z > 20$ the value of $\Delta_t$ changes weakly as a function of maximum redshift. Right panel: probability distribution in arbitrary units for $\Delta_t$ having marginalised over $H_0$, $\Omega_{m,0}$ and $z_f$ (see text for details). This shows that the current range for $\Delta_t$ is small and smaller than current uncertainties in the stellar ages of the oldest stars. The left side of the PDF is sensitive to $z_f$ but very weakly as it falls steeply. The right side of the PDF is sensitive to the redshift of  the farthest known galaxy.}  
\label{fig:deltat}
\end{centering}
\end{figure}

Local, precise determinations of ages of the oldest stars or stellar populations $t_{\rm oldest\; stars}$ can be used to estimate the age of the Universe, $t_U$, in a way that is very weakly sensitive to the adopted  cosmological model. The maximum age of stellar objects is just a lower limit to the age of the Universe, $t_U = t_{\rm oldest\;stars} + \Delta_t$, where  $\Delta_t$ is  taken to be the age between the Big-Bang and the formation of the oldest stars observed. To derive a value for $\Delta_t$ we proceed in the following way. Galaxies are known to exists at $z\gtrsim 11$ \cite{Oesch} and  high-redshift galaxies in larger numbers are found and confirmed at $z > 8$.  It is therefore reasonable to assume\footnote{If the Universe is homogeneous and isotropic, then our Galaxy must be a typical object (Copernican principle). Hence the oldest, extremely low metallicity stars in our galaxy must belong to the typical population of the most distant  and thus oldest objects in the Universe, and these are observed at redshifts well above $z=8$. }  that the formation redshift $z_f$ for the oldest stars observed is $z_f >8$.   Here we assume $z_f\ge11$. Assuming $z_f>8$ does not change the results significantly, see below. These objects  cannot have formed  before  reionization, which is believed to have started around $z\sim 30$, however, as we will show below, this limit is unimportant. We can compute then $\Delta_t$  as the time between $z= 11$ and $z_{\rm max}$, which depends very weakly on cosmology  for reasonable choices (despite the $H_0$ tension) and very weakly on  $z_{\rm max}$ provided $z_{\rm max}\gtrsim 20$. This is shown in the left panel of  Fig.~\ref{fig:deltat}, where we plot the value of $\Delta_t$ as a function of the redshift of formation of the oldest stars. From this figure one can see that  a conservative estimate is $\Delta_t \sim 0.5-0.1$ Gyr for the values of $\Omega_{m,0}$ and redshift of interest.  Note that identifying the oldest observed stars  as the descendant of the lower-redshift galaxy observed (i.e.,  setting $z_f>8$) would increase $\Delta_t$ by $\sim 0.1$ Gyr.

\subsection{Inferring $H_0$ from $t_{\rm oldest\,stars}$ in a (quasi)$\Lambda$CDM model}
\label{sec:methods:H0}
Today, the standard cosmological model is spatially flat, but cosmological constant-dominated; there is strong evidence and strong theoretical prejudice for a spatially flat Universe, in which case  age, $t$, and $H_0$ are related by:
\begin{equation}
H_0 = \frac{A}{t} \int_0^{z_t}  \frac{1}{1+z} \left [  \Omega_{m,0} (1+z)^3 + (1-\Omega_{m,0}) (1+z)^{3 (1+w)} \right ]^{-1/2} dz
\label{eq:h0}
\end{equation}
where $A=978$ if $H_0$ is in km s$^{-1}$ Mpc$^{-1}$, and $t$ is the time  from today until $z_t$ in Gyr.  The age of the Universe $t_U$  corresponds to  setting $z_t = \infty$, although for most reasonable cosmological models by $z$ $\sim$ a few tens the integral has effectively converged. 
  
We have ignored both the radiation and curvature contributions since they are robustly constrained, independently of the $\Lambda$CDM model, to be negligible compared to the matter and dark energy densities. Note that we  do not specify  the value of the equation of state parameter of dark energy $w$ (assumed constant), although the $\Lambda$CDM model assumes $w=-1$. 

While $t_U=t_{\rm oldest\,stars}+ \Delta_t$,
 to estimate $H_0$ from a  determination of $t_{\rm oldest\,stars}$  it is not necessary to  estimate $\Delta_t$ and pass through $t_U$, we  just use Eq.~  \ref{eq:h0} for $t=t_{\rm oldest\,stars}$. There will be an uncertainty on $z_t$, which in turn depends (weakly) on cosmology, which can be propagated into error-bars. Since we use standard Markov Chain Monte Carlo to sample to posterior distribution of parameters this is done straightforwardly. 
 
 \subsection{Constraining $\Omega_{m,0}$ (and $w$) from the low $z$ Universe}
 \label{sec:methods:Omm}
 
The local density parameter of the Universe can be determined  very easily and precisely  from CMB observations, but for our purposes we need a late-time only measurement, independent from CMB physics. 
Precise measurements of the expansion history via BAO in the galaxy distribution and SNe constrain the combination $F_{AP}(z) \equiv D_A(z)H(z)$ (Alcock-\paczynski\ test, where $D_A$ is the angular diameter distance) which does not depend on $H_0$ nor on the sound horizon at radiation drag, and of $D_L$, the luminosity distance, respectively. These quantities however depend on $\Omega_{m,0}$ (and $w$). Redshift space distortions probe the growth of structure hence the linear growth of perturbations, which, assuming general relativity, is related to $\Omega_{m,0}$.  Higher order correlations can help break degeneracies between linear growth rate and other quantities such as bias and dark matter clustering amplitude {\cite{Verde2dF,Bis1,Bis2,Bis3} for models with a nearly scale invariant power spectrum.  Another probe is the Alcock-\paczynski\ test with cosmic voids \cite{voidAP}. While it was suggested relatively recently, its power has been demonstrated successfully on simulations and data, and refinements of the method are being readied for next generation surveys (e.g.,~\cite{voids,Nadathur} and references therein). All these methods  assume a $\Lambda$CDM-type cosmology  but are independent of the CMB and do not rely on assumptions about early time physics. 

\subsection{Consistency checks: a proof of principle approach}
The age of the Universe estimated  as in Sec.~\ref{sec:methods:ages} can be  directly compared with that inferred from CMB observations. This provides a long lever-arm consistency check: $t_U$ extrapolated from observations at $z=1100$ can be compared with that estimated at $z=0$.

On the other hand, for a flat $\Lambda$CDM model, Eq.~\ref{eq:h0},  relates together three quantities $\{ H_0, t_U, \Omega_{m,0} \}$, hence the value of two of them determines the value of the third. These  three quantities can be all determined or measured either at low redshift thus relying on late times physics, as in Ref.~\cite{Riess+18} and Sec.\ref{sec:methods:ages}--\ref{sec:methods:Omm},  or  extrapolated from  CMB  data (relying on early time physics).  This offers a powerful consistency check. 

For a flat dark energy dominated CDM model (wCDM) where the equation of state parameter of dark energy is  constant but not necessarily $w=-1$,  Eq.~\ref{eq:h0} presents  an additional variable and therefore there is an extra degree of freedom.  Nevertheless for  invoking a $w\ne -1$ as a solution to the low vs high redshift tension  all the data sets must be consistent for the same value of $w$. 
   
Here we present a proof of principle of such a consistency check using current data.
While current  determinations of $t_U$ from the local Universe might not have the robustness and accuracy to match those of local $H_0$ determinations, forthcoming data will greatly improve the situation.

\section{Data}
\label{sec:data}
For our proof-of-principle application, the data we consider are:
\begin{itemize}
\item The local determination of the Hubble constant \cite{Riess+18} using the distance ladder from Cepheids. Their reported value is $H_0 = 73.48 \pm 1.66$ km s$^{-1}$ Mpc$^{-1}$.

\item The Planck 2018 \cite{Planck18} data (Temperature, Polarisation and CMB lensing signals) analysed with standard pre-recombination physics, for a $\Lambda$CDM model and a $w$CDM model. The posterior samples for the relevant  parameters are publicly available trough the Planck Legacy Archive. In a flat $\Lambda$CDM modle Planck yields $H_0 = 67.37 \pm 0.54$ km s$^{-1}$ Mpc$^{-1}$ and $t_U=13.8 \pm 0.02$ Gyr.

\item Absolute ages of galactic globular clusters as derived by Ref.~\cite{omalley}.  The authors computed ages for 22 galactic globular clusters using the latest Gaia distances for sub-dwarfs to calibrate the distance to the globular clusters. One important feature of their analysis is the  very careful and thorough effort to include all systematic effects in the uncertainty of the age determination. In particular, they included all theoretical uncertainties in the modeling of stellar physics, that in turn is used to compute the isochrones to derive the luminosity of the main sequence turn-off. Further, they paid special attention to fitting individual chemical abundances,  in particular  to the presence of alpha-enhanced elements, and  obtained a distance determination to each GC using updated Gaia distances to local sub-dwarfs. We adopt the absolute stellar ages derived by the authors in their Table~6 using the luminosity of the main-sequence turn-off for each GC individual color-magnitude diagram. We  selected globular clusters with metallicities below $[{\rm Fe/H}] < - 2$ as they have been shown to be coeval and also the oldest.  A weighted average  yields $t_{\rm GC}=13.0 \pm 0.4$ Gyr as an estimate\footnote{Constraining the oldest age by averaging the individual age estimates assumes that the   $[{\rm Fe/H}]$ selection makes the intrinsic scatter in the ages of these GCs small compared to their error bars.}
 of $t_{\rm oldest\;stars}$.

\item The absolute ages of very-low-metallicity stars. In particular, there are two stars we are aware of with $[{\rm Fe/H}] < -2 $ that have reliable ages and Gaia parallaxes. The stars we considered are: HD140283~\cite{Bond} and J18082002-5104378 A~\cite{Schlaufman}. The values we adopt are $13.5 \pm 0.7$ Gyr for HD140283 and $13.0 \pm 0.6$ Gyr for J18082002-5104378. J18082002-5104378 is a very-low-metallicity star ($[{\rm Fe/H}] = -3.5$) in a binary system. Ref.~\cite{Schlaufman} report a value for the age of $13.535 \pm 0.002$ Gyr. This corresponds to the best fitting isochrone that includes $\alpha-$enhancement. The exceedingly accurate fit is not surprising when using Bayesian techniques to estimate the age~\cite{Feltzing} while marginalizing on the other parameters; similar small formal uncertainties are obtained in Ref.~\cite{Ata} when fitting nearly 70 GCs with Bayesian techniques and the use of a single stellar model. Ref.~\cite{Schlaufman} also fit  two other stellar models and find ages for the stars that are $1-2$ Gyr lower; however, these isochrones have no $\alpha-$enhancement; the fit with $\alpha-$enhancement has a much better goodness of fit.  Because of the  widely different physical assumptions, the dispersion among these three different age determinations offers an estimate of possible systematic errors.  The  Bayesian technique \emph{Baccus}~\cite{Baccus} is particularly suited to this application as by construction it provides joint probability distributions  for multiple measurements assuming they are all affected by unknown systematic shifts. We report the mean and  68\% confidence range\footnote{But note that \emph{Baccus} distribution has tails that are wider than Gaussian.} between the three different age determination obtained with  \emph{Baccus}. For HD140283 we have adopted the new Gaia DR2 parallax of $16.1 \pm 0.07$ mas (twice more accurate than the HST parallax) which lowers the age estimated in \cite{Bond} using HST parallaxes by 1 Gyr.\footnote{We have checked and estimated uncertainties for these stars and the GCs above using the very useful Python package \emph{isochrones}~\cite{isochrones}. See also the plots in Ref.~\cite{JimenezMac} for intuition on the effect of physical parameters on the stellar tracks.}

\item The low-redshift matter density parameter determination in a flat Universe can be obtained in different ways. 
A pioneer measurement, using a combined analysis of power spectrum and  higher-order correlations,  was presented for the Two-degree Field Galaxy Redshift Survey in Ref.~\cite{Verde2dF}: $\Omega_{m,0}=0.27 \pm 0.06$.
A combined analysis of power spectrum and  higher-order statistics of $z\sim 0.45$ galaxies   (see Ref.~\cite{Bis1,Bis2,Bis3}) find a value $\Omega_{m,0} = 0.341 \pm 0.045$; this measurement  uses bispectrum measurements and redshift space distortions for the BOSS galaxy catalog, assuming a $\Lambda$CDM-like model. This is a measurement independent of  the CMB.  The low redshift matter density parameter has also  be obtained by the Alcock-\paczynski\ test applied to voids (also for a $\Lambda$CDM-like model) in  Ref.~\cite{voids}: $\Omega_{m,0} = 0.28 \pm 0.03$. This value is  also derived independently of the CMB. The three  $\Omega_{m,0}$ measurements are fully consistent, their combination yields $\Omega_{m,0}= 0.3\pm 0.023$.
 Finally the Alcock-\paczynski\ test applied to BOSS galaxies yields $\Omega_m=0.290\pm 0.053 $ but for a $w$CDM model.
  This determination is in excellent agreement with that obtained in a $\Lambda$CDM model with the Pantheon supernova sample \cite{Pantheon} $\Omega_{m,0}=0.298 \pm 0.022$, which  in a $w$CDM model  becomes $\Omega_{m,0}=0.316 \pm 0.072$. In summary we adopt a prior of  $\Omega_{m,0}= 0.3\pm 0.02$ which we assume Gaussianly distributed,  for a $\Lambda$CDM Universe.
\end{itemize}
 
 To obtain $t_U$ from $t_{GC}$  and $t_*$, we sample the $\Delta_t$ distribution via MCMC;  we impose a  Gaussian prior on $H_0$ of $70\pm 10$km s$^{-1}$ Mpc$^{-1}$, a  generous Gaussian prior on $\Omega_{m,0}$ of $0.3\pm 0.1$ and a log prior on $z_f$ between $z=11$ and $z=30$ although the upper limit does not matter.  We obtain a  skewed distribution of $\Delta_t$ (shown in the right panel of  Fig~\ref{fig:deltat}) with a median of $0.2$ Gyr, a most probable value at $0.12$ Gy and 68\% (95\%) confidence $0.08<\Delta_t<0.26$ ($0.065<\Delta_t<0.45$).
 
 Very similar results are obtained for uniform priors on $H_0$ and $\Omega_{m,0}$. For the purposes of plots, since the additional error induced on $t_U$ by the  
 $\Delta_t$ uncertainty is not dominant, we ignore the small non-Gaussian behavior of  $\Delta_t$ and  use instead an effective  Gaussian 
 $\Delta_t=0.2\pm 0.1$.  This is shown in Fig. \ref{fig:ages} where the dotted line is obtained from $t_{\rm CG}$ as determined by Ref.~\cite{omalley} ($t_U^{GC}=13.2\pm 0.4$); the  black dashed line corresponds to $t_*$  estimated using the age of the very-low-metallicity stars  ($t_U^{*}=13.4\pm 0.45$). Assuming $z_f>8$ would increase the derived $t_U$ by 0.1 Gyr; for current error-bars this  would correspond to a shift of less than  $\sim 0.3 \sigma$.  For comparison, the solid line is the $\Lambda$CDM-model-dependent value derived by Planck18.
 All effects  due to the skewness in the $\Delta_t$ distribution are at the level of  less than one tick mark in the plot and hence  do not affect qualitatively our conclusions\footnote{This approximation does not affect the $H_0$ inference as $\Delta_t$ does not enter in that calculation} . At this point we prefer not to combine the two  stellar estimates in inferring $t_U$ nor  to constrain $H_0$. In the inset, and just as an example of what kind of accuracy could be obtained if systematic uncertainties were all under control, we plot the age for the formal uncertainty of  J18082002-5104378, which is fully compatible with Planck18.
 
\begin{figure}
\begin{centering}
\includegraphics[width=.7\columnwidth]{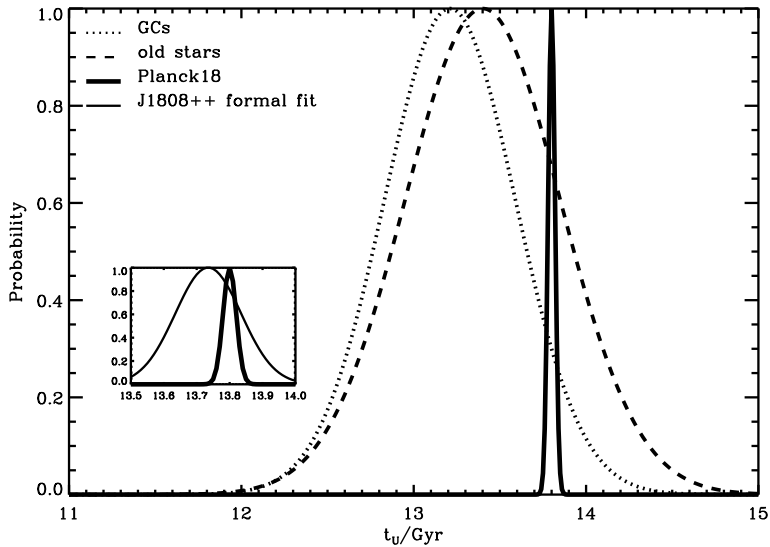}
\caption{Probability distribution for the age of the Universe $t_U$ obtained using stellar ages and that derived by Planck18 from the CMB assuming the $\Lambda$CDM model. There is good agreement between both stellar ages and the Planck18 derived value. In the inset, and just as an example of what kind of accuracy could be obtained if systematic uncertainties were all under control, we plot the age for the formal uncertainty of  J18082002-5104378, which is fully compatible with Planck18. The formal GCs ages for 69 ACS clusters from Ref.~\cite{Ata} would look similar to the J18082002-5104378 line.} 
\label{fig:ages}
\end{centering}
\end{figure}

\section{Results}

Stellar ages provide an estimate of $t_U$ compatible with the value obtained by Planck18 assuming the $\Lambda$CDM model. The ages of GCs are slightly lower, but at less than the $1\sigma$ level, thus  the difference is statistically insignificant.
We now consider  more closely the implication of this $t_U$ estimate in light of our selected data set  and in relation to Eq.~\ref{eq:h0}.
This is illustrated in Fig.~\ref{fig:local} where we report the constraints in the $t_U$-$H_0$ plane.

\begin{figure}
 \begin{centering}
\includegraphics[width=.48\columnwidth]{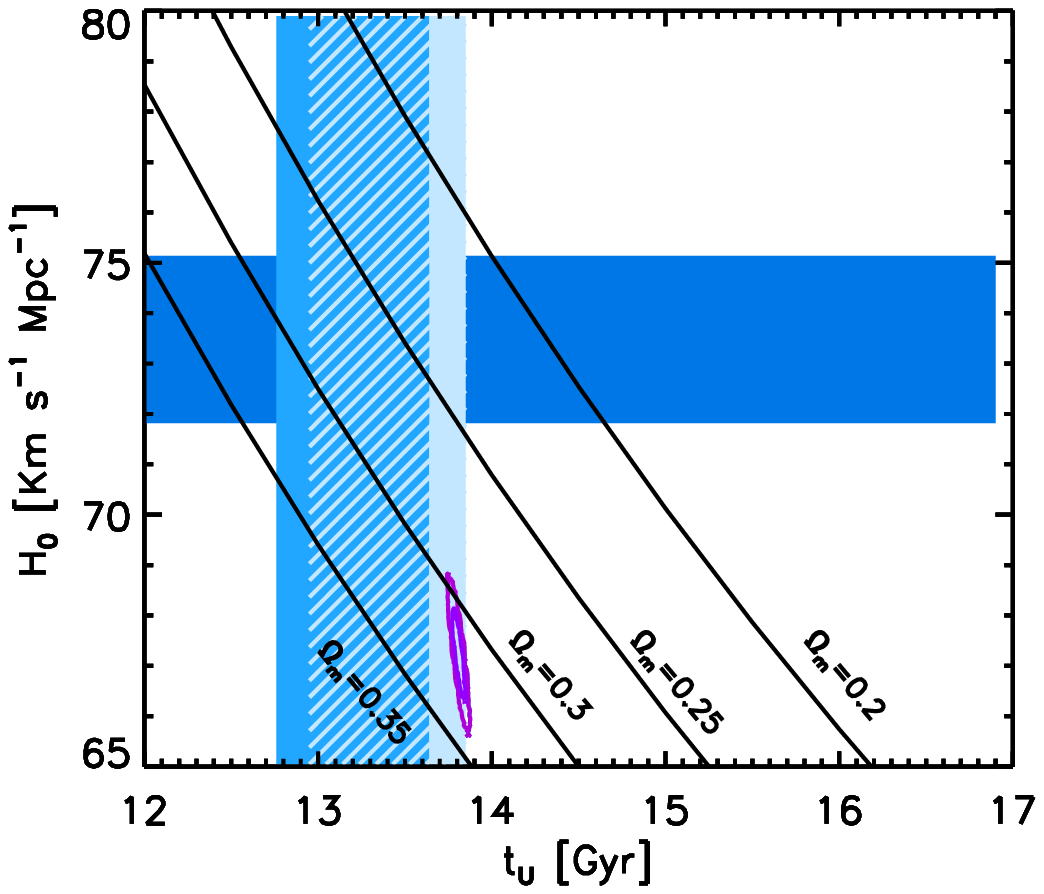}
\includegraphics[width=.48\columnwidth]{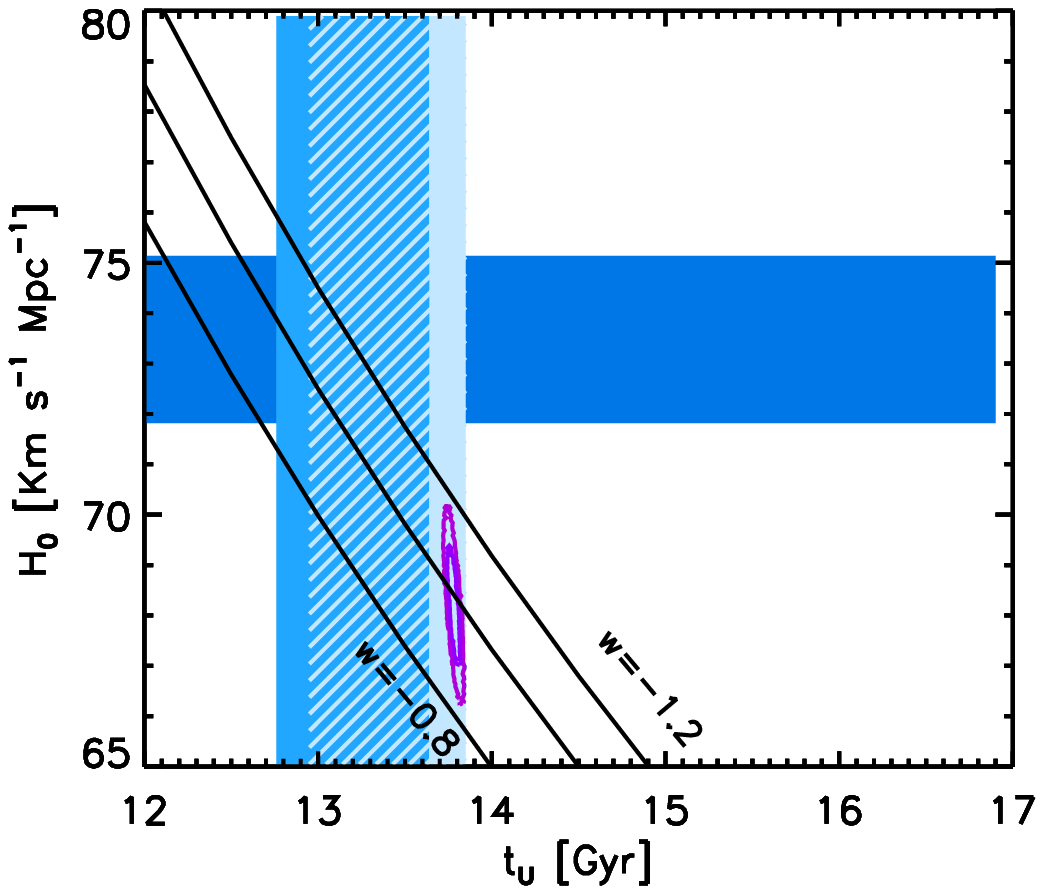}
\caption{$t_U$ as obtained from stellar ages (vertical light blue bands: GC darker band, ultra-low-metallicity star lighter band, patterns where the two ranges overlap) vs. $H_0$  local distance ladder determination (horizontal dark blue band); $1-\sigma$ uncertainties are shown (see text for more  details about uncertainties). Also plotted is the CMB-predicted constraints  from  Planck18 (magenta contours, 68 and 95\% confidence level shown) in a flat $\Lambda$CDM model in the left panel and for  Planck18+BAO and SNe and  wCDM model  in the  right panel. We also plot lines of constant $\Omega_{m,0}$ (left panel, $w=-1$) and $w$ (right panel,  $\Omega_{m,0}=0.3$). The plot illustrates that changing the value of $\Omega_{m,0}$ or $w$ does not alleviate the local vs. early Universe tension (see text). But it is clear that $t_U$ determinations, along with a  $\Omega_{m,0}$ prior, could offer precious information to understand the origins of this tension, once their error-bars shrink even only moderately.}    \label{fig:local}
\end{centering}
\end{figure}  

The light blue bands are estimates of $t_U$ using stellar ages as described above (the darker one centered around  13.2 Gyr is from GCs ages, while the lighter  one centered around 13.4 Gyr is  from the very-low-metallicity stars), while the $H_0$ value, horizontal dark blue band, is the one determined by the local distance ladder \cite{Riess+18}. All bands are $1\sigma$ confidence levels. 
The magenta contours are the ones from Planck18 assuming a $\Lambda$CDM model (left panel) and from Planck18+BAO+SNe assuming a $w$CDM model (right panel).  Note that by adding SNe and BAO to CMB, we implicitly calibrate the  ``inverse" distance  ladder to CMB.  For CMB only data the $H_0$-$t_U$ degeneracy extends almost vertically in the plot,  to high $H_0$ values and  $w<-1$, leaving the age almost unchanged. These models with such negative $w$ values however, are disfavoured by BAO and  especially by cosmological SNe observations.  We also show lines of constant $\Omega_{m,0}$ (for $w=-1$) on the left panel   and  constant equation of state of dark energy (for $\Omega_{m,0}=0.3$) on the right panel. 
This figure  illustrates further the tension: no choice of  $\Omega_{m,0}$ or $w$
can bring into agreement the three data sets. But  $t_U$ determinations, along with a  $\Omega_{m,0}$ prior, could offer precious information to understand the origins of this tension, once their error-bars shrink even only moderately.

While these considerations are mostly qualitative because in details they depend on an estimate of  the added age between the Big-Bang and the formation of the stars ($\Delta_t$), it is possible to simply treat $t_{\rm GC}$ or $t_*$ as a lower limit to $t_U$.  In this case,  for example, should a $t_{\rm oldest\;stars}$ be found to be reliably  at least $13.6$ Gy old,  and $\Omega_{m,0}>0.25$ from low redshift Universe observations only, it would  support a CMB-calibrated $\Lambda$CDM model and would confine the solution to the ``Hubble-tension" to very local physics.

We can now use Eq.~\ref{eq:h0}  to determine $H_0$  within a $\Lambda$CDM model using only late-time Universe  observations. 
 In doing so we  set $t$ to be $t_{\rm CG}$ or $t_*$, and Eq.~\ref{eq:h0} is sampled via standard MCMC. A logarithmic prior on $z_t$ is imposed for $11< z_t<30$ although the result is insensitive to the upper limit of this range. We also impose a Gaussian prior on $\Omega_{0,m}$ as in Sec.~\ref{sec:data}. 
Fig.~\ref{fig:h} shows the  resulting probability distribution for $H_0$, using $t_{\rm CG}$ on the left panel and using $t_*$  on the right.  For comparison, we display the $1\sigma$ ranges from Planck18 (for a $\Lambda$CDM model) and the local  distance ladder  method \cite{Riess+18}. GCs ages imply a  best fitting value of $H_0= 71.0 \pm 2.8$ km s$^{-1}$ Mpc$^{-1}$ and  the very-low-metallicity stars yield $H_0= 69.3 \pm 2.7$ km s$^{-1}$ Mpc$^{-1}$.  Assuming $z_f>8$ would yield  $H_0$ estimates lower by 0.2  km s$^{-1}$ Mpc $^{-1}$; this shift is negligible compared to the statistical error-bars, confirming that the main source of uncertainty at the moment is not $t_{\rm oldest\;stars}$ and  their related $z_f$.

To consider deviations from the standard $\Lambda$CDM model in the form of a $w$CDM cosmology, an additional prior in $w$ must be introduced to constrain this extra degree of freedom.  The dark energy equation of state parameter can be constrained, in a $w$CDM cosmology,  from late-Universe  observations only, independently from the CMB. In fact $w$  effectively changes the behaviour of the Universe expansion history at relatively low redshifts, where dark energy dominates. For example the Pantheon supernova sample \cite{Pantheon} constrains $w=-1.09\pm0.22$.
When including this prior, for a $w$CDM  model, the inferred $H_0$ becomes $H_0=71.8 \pm 3.6$ for CGs and  $H_0= 71.0 \pm 2.8$ for the star, this is shown as dashed lines in Fig.~\ref{fig:h}.

\begin{figure}
\begin{centering}
\includegraphics[width=.7\columnwidth]{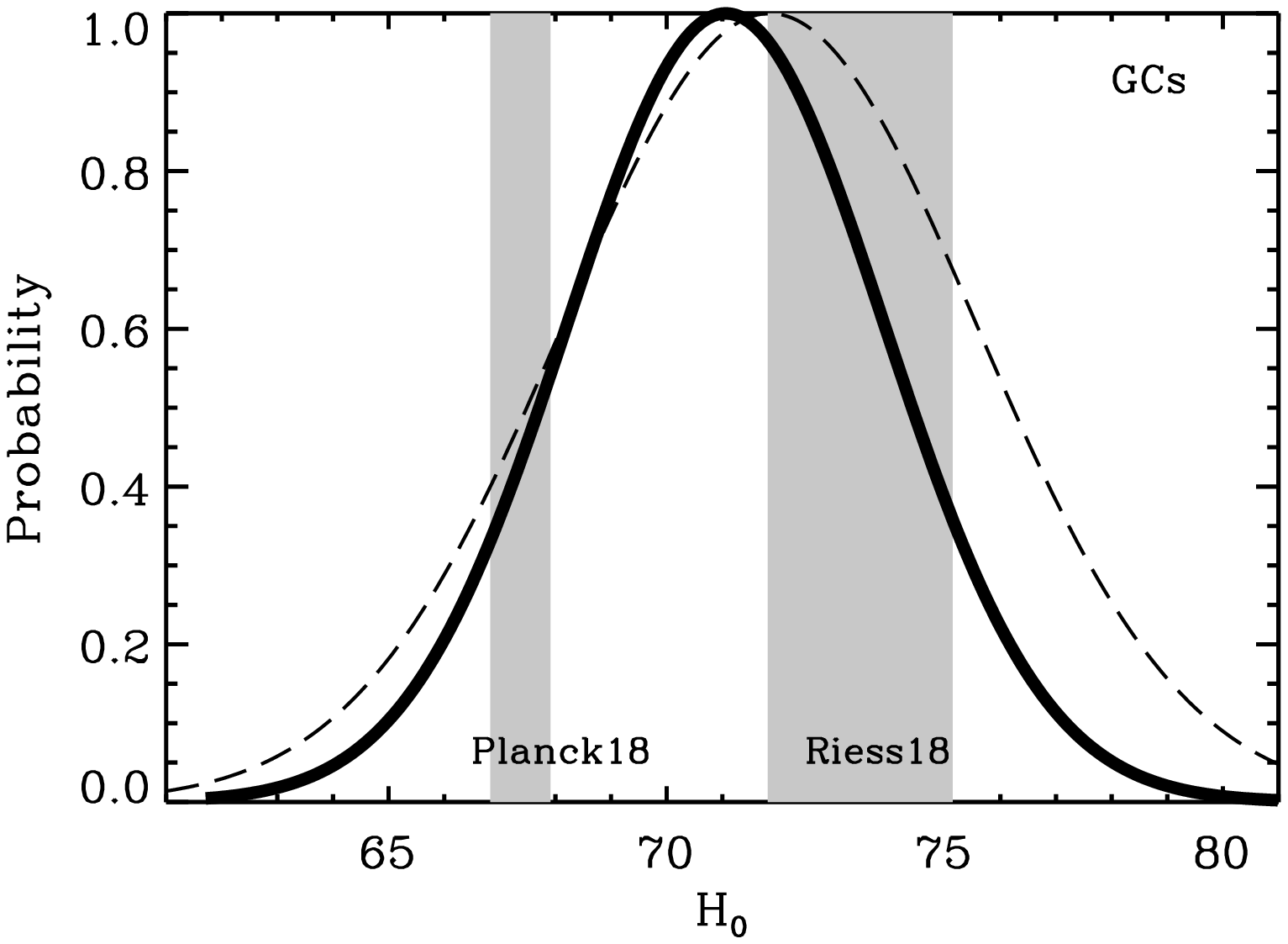}
\includegraphics[width=.7\columnwidth]{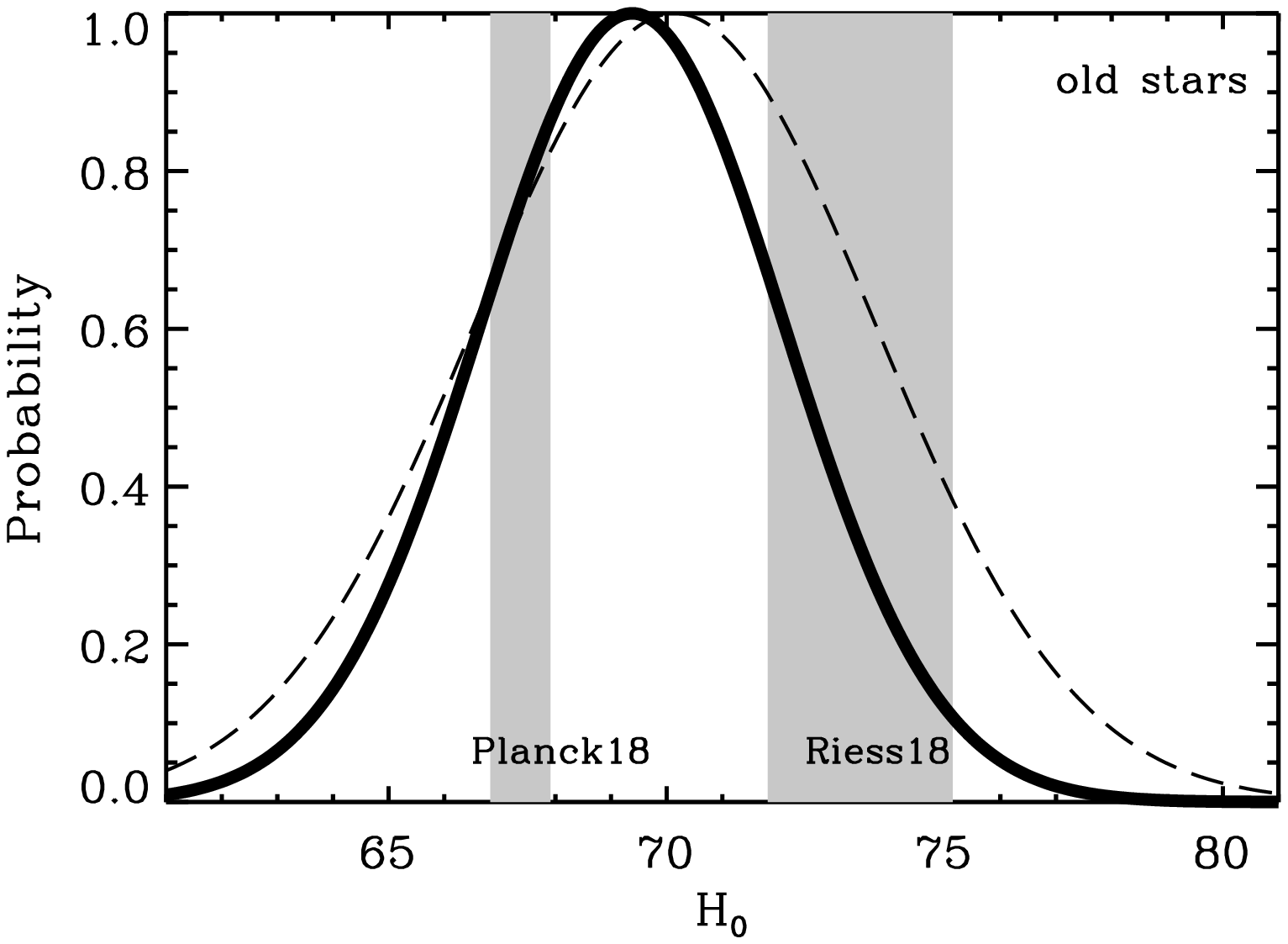}
\caption{Probability distribution (arbitrary normalisation) for $H_0$ using stellar ages and the local value of $\Omega_{m,0}$ (see text). Upper panel: ages from the globular cluster sample~\cite{omalley}. The best fit value is $H_0= 71.0 \pm 2.8$ km s$^{-1}$ Mpc$^{-1}$. Lower panel: stellar ages from the very-low metallicity stars. The best fit value is $H_0= 69.3 \pm 2.7$ km s$^{-1}$ Mpc$^{-1}$. The grey shaded areas show the allowed 1$\sigma$ interval for the cosmological-model dependent $\Lambda$CDM-based $H_0$ derived by Planck18 and the cosmology-model-independent cosmic distance ladder value derived by Ref.~\cite{Riess+18}. Dashed lines show the $H_0$ constraints obtained in a $w$CDM model assuming a $w$ prior given by the Pantheon supernovae sample \cite{Pantheon}.}
\label{fig:h}
\end{centering}
\end{figure}

\begin{table}
\begin{center}
\caption{Values for $H_0$ obtained by different methods, in units of km s$^{-1}$ Mpc$^{-1}$: Planck18, derived value assuming $\Lambda$CDM \cite{Planck18}; Local ladder, from Cepheids\cite{Riess+18}; GW, electromagnetic counterpart to gravitational waves\cite{GW}; CC14 and CC18, cosmic chronometers using Gaussian-Process methods\cite{GP,morescoGP}; GCs and low-met stars are the values obtained in this work using stellar ages and CMB independent inferences of $\Omega_{m,0}$.}
\label{tab:h0}
\vspace*{0.5cm}
\begin{tabular}{ccc}
\hline
method & $H_0$ & uncertainty \\
\hline
Planck18 & 67.37 & 0.54 \\
Local  ladder & 73.48 & 1.66 \\
GW & 70 & +12-8 \\
CC14 & 68.5 & 3.5 \\
CC18 & 68.5 & 2.9 \\
GCs & 71.0 & 2.8 \\
low-met stars & 69.3 & 2.7 \\
\hline
\end{tabular}
\end{center}
\end{table}

\section{Discussion}

While current age measurements  do not yet have the statistical power to shed  light on the  early vs late Universe tension,  even a modest reduction of the errors will.
It is  thus useful to consider how  statistical and systematic errors can be reduced with future data and  better theoretical modeling.

Systematic uncertainties in the age of stars  are  related to  the stellar modeling.  Currently, the  main systematic uncertainties  include: nuclear reaction rates; element opacities; physics of the convective envelope and primordial Helium abundance. For globular clusters ages there is an additional uncertainty  associated to the luminosity of the main sequence turn-off.

The first three sources of errors can in principle  be mitigated with better theoretical modeling.  There is no fundamental  limitation to improving  models of stellar convection, which still relies on 1D modeling. This could be done  for example by  sampling the space  of relevant variables  with  3D detailed stellar models (which are more accurate but more computationally demanding), and then use these to calibrate the 1D models. Likewise,  calculation of nuclear reaction, and related opacities,  rates could in principle be  improved  with more computational effort (time and resources).

The primordial Helium abundance is often obtained from CMB observations (hence making the age determination not fully CMB independent). However, the CMB-derived value is fully consistent with determinations from astronomical objects \cite{Aver}, see e.g., Fig.~40 in Ref.~\cite{Planck18}. 

The current biggest observational uncertainty on estimating the luminosity of the main sequence turn-off is the distance to the GCs. It is expected that this will be  dramatically reduced with direct Gaia parallaxes to the oldest GCs.

In summary, the main uncertainties in deriving stellar ages are distances and abundances and it is here where a coordinated effort (modelers and observers) should take place.

There is a residual additional error  introduced mainly by the lower limit  in the uncertainty in $z_f$ the formation redshift of the stars. Since the farthest know galaxy found today is at $z=11.1$,  we have assumed that this corresponds to the lower limit for $z_f$. Should galaxies be found  at higher redshifts, this argument would  directly reduce both  the $z_f$-induced error on $H_0$ and that in $\Delta_t$.

Having said this, the current uncertainty in the  $\Omega_{m,0}$ determination from the low-redshift Universe is now the dominant error component for the low-metallicity star $H_0$ determination.  Hence a drastic reduction in GC or stellar ages at the moment would not change drastically our results. However, forthcoming galaxy redshift surveys are  expected  to improve dramatically the $\Omega_{m,0}$ constraint used here.

It is interesting to speculate on what accuracy on the age of stellar objects is needed to securely (at the 95\%) rule out or in either value of $H_0$  (Planck18 or the local-distance ladder). Let us assume that in the near future  $\Omega_{m,0}$ is measured with half the current uncertainty ($0.01$) which is reasonably  conservative considering the volume covered by  forthcoming  and future large scale structure surveys. Then for several fiducial values of $t_{\rm oldest\, stars} $ we can use Eq.~\ref{eq:h0} to forecast the needed accuracy. We find a value of $0.3$ Gyr for \emph{an individual object}. Clearly the statistical error is not a problem as even with current uncertainties of $\sim 1$ Gyr, just ten old stars or GCs would be enough to achieve this. What is needed is to control systematic uncertainties better than $\sim$ few percent.  The final Gaia analysis will provide {\em direct} distances to GCs (up to 15kpc) at  percent accuracy \cite{Gaiasim}. 

The inference of stellar ages from stellar spectral energy distributions or color-magnitude diagrams has come a long way since the early pioneering days. The future looks bright for careful treatments of age inferences for the oldest stars in our Galaxy in the spirit of Ref.~\cite{Ata,Feltzing} that carefully include Bayesian marginalization over nuisance parameters

\section{Conclusions}

Stellar ages  can provide a powerful test of  cosmology based on local Universe observations. Historically, before the direct evidence for the accelerating Universe from SNe at the end of the 1990-ies, stellar ages pointed to the ``older than the Universe" problem \cite{JimenezGC,ostriker,Jaffe,Primack,Chaboyer},  and  did ``guide one forcefully to examine models" beyond the  standard model of cosmology of the time: the Einstein-deSitter model~\cite{ostriker}. The accuracy of stellar ages has improved tremendously in the last 20 years, especially in the reduction of the systematic uncertainty; interestingly enough, the central value for the estimated age of the oldest stellar objects has stayed nearly identical (see e.g. Ref.~\cite{JimenezGC}). 

While it is hard to overstate how much  observations of the CMB have been fundamental to establish a new standard model of cosmology,
the importance of local measures in cosmology has recently become clear e.g.,  \cite{Suyu2012, VerdeJimenezFeeney, 2017NatAs, Heavensruler, Verderuler}. Purely  late-time measurements of cosmological parameters, when compared to CMB-derived constraints,  allow one to perform ``end to end" tests of our understanding of the Universe and the physics underlying the standard cosmological model.

This has attracted renewed interest in light of the $\sim 3\, \sigma$ tension between early Universe (CMB-based) and late-Universe determinations of $H_0$. A confirmation  with independent data sets and methodology, that the $H_0$ tension is indeed a late-vs early time issue, would inform us on what  new physics should be invoked to resolve it.

The cosmologically relevant quantities that can be measured from the low-redshift Universe are not limited to $H_0$, these are: $H_0$ (through several different and independent approaches), $t_U$ and $\Omega_{m,0}$. In a $\Lambda$CDM Universe these three quantities are related  hence these measurements can be used as a powerful consistency check of the model. 

We have presented a proof-of-principle  application of this approach on current data (see Table~\ref{tab:h0}). While with current error bars on $\Omega_{m,0}$ and the age of the oldest stars it is not possible to decide between Planck18 or the local distance ladder values, we note that a simple factor two reduction of current uncertainties would 
provide already a conclusive test.
 
As uncertainties on ages and on  late-Universe determinations of $\Omega_{m,0}$ shrink,  we can foresee several different outcomes.
\begin{itemize}
\item[{\it i)}] GC-derived ages and low-met stars-derived ages do not agree with each other.  Since the modelling relies on isochrones and these are the same for both objects this would indicate a systematic in  the distance determination. But if this were to imply a systematic in the parallaxes, it would have profound consequences for other astronomical applications too.

There is however at the moment no reason to believe that this will happen. 
\item[{\it ii)}] Stellar ages-derived $H_0$ determinations are consistent with CMB-inferred $H_0$ values in a $\Lambda$CDM model. This may indicate that a solution to the $H_0$ tension must be looked for in very local, nearby physics.
\item[{\it iii)}] Stellar ages-derived $H_0$ determinations are consistent with local distance ladder measurements. This will give an indication  that  new physics must be sought in the two decades of expansion history  before recombination. This motivation  would be even more compelling if  standard sirens and time delay distances  also agree with the local measurements. 
\item[{\it iv)}] Stellar ages-derived $H_0$ determinations fall in between and exclude both.  This would make us reconsider the $H_0$ problem as an anchor problem and maybe look into  physics that can introduce suitable  modifications of the expansion history of the Universe (and yet be consistent with a host of observations such as BAO and SNe).
\item[{\it v)}] The nightmare scenario: the four different late-Universe  $H_0$ determinations (local distance ladder, stellar ages-derived, time-delay distances and standard sirens) do not agree with each other and do not resolve the tension with the early Universe. The physics to explain this case would likely  be highly  contrived, thus this would indicate  widespread and unknown systematic issues affecting different measurements well above expectations.
\end{itemize}
Scenarios {\it ii)} and {\it iii)} would be the most interesting and are probably the most reasonable to expect. In this case, the future release of Gaia accurate and robust {\em direct} distances to the oldest GCs will provide us with ages of these stellar object accurate enough as to    determine if a tension is also present in the age of the Universe,  confirm whether radically  different aspects of the low redshift Universe  consistently prefer higher $H_0$ values  and  help sift among possible solutions and explanations  to the Hubble tension if it remains.

\[ \]{\bf Acknowledgments:} We thank Adam Riess and George Efstathiou for very useful comments and Jose Luis Bernal for providing us with the Baccus estimates. RJ and LV thank the Center Emile Borel for hospitality during the latest stages of this work. Funding for this work was partially provided by the Spanish MINECO under projects AYA2014-
58747-P AEI/FEDER, UE, and MDM-2014-0369 of ICCUB (Unidad de Excelencia Mar\'ia de
Maeztu).  LV acknowledges support by European Union's Horizon 2020 research and innovation program
ERC (BePreSySe, grant agreement 725327). AC and MM acknowledge the grants ASI n.I/023/12/0, PRIN MIUR 2015 and ASI n.2018-23-HH.0.  BDW acknowledges support by the Labex Institut Lagrange de Paris (ILP) (reference ANR-10-LABX-63) part of the Idex SUPER, and received financial state aid managed by
the Agence Nationale de la Recherche, as part of the programme Investissements d'avenir
under the reference ANR-11-IDEX-0004-02. This work was supported by the Simons Foundation. BDW is supported by the ANR BIG4 grant ANR-16-CE23-0002 of the French Agence Nationale de la Recherche.

\end{document}